\documentclass[prl,twocolumn,aps,superscriptaddress,showpacs]{revtex4-1}

\usepackage{amsmath,amssymb}
\usepackage{graphicx}
\usepackage{xcolor}
\usepackage[final]{changes}

\begin{document}

\title{\textbf{Lifshitz Transitions in the Ferromagnetic Superconductor UCoGe}}

\author{Ga\"el Bastien}
\email{gael.bastien@cea.fr}
\affiliation{University Grenoble Alpes, INAC-PHELIQS, F-38000 Grenoble, France}
\affiliation{CEA, INAC-PHELIQS, F-38000 Grenoble, France}
\author{Adrien Gourgout}
\affiliation{University Grenoble Alpes, INAC-PHELIQS, F-38000 Grenoble, France}
\affiliation{CEA, INAC-PHELIQS, F-38000 Grenoble, France}
\author{Dai Aoki}
\affiliation{University Grenoble Alpes, INAC-PHELIQS, F-38000 Grenoble, France}
\affiliation{CEA, INAC-PHELIQS, F-38000 Grenoble, France}
\affiliation{IMR, Tohoku University, Oarai, Ibaraki 311-1313, Japan}
\author{Alexandre Pourret}
\affiliation{University Grenoble Alpes, INAC-PHELIQS, F-38000 Grenoble, France}
\affiliation{CEA, INAC-PHELIQS, F-38000 Grenoble, France}
\author{Ilya Sheikin}
\affiliation{LNCMI-EMFL, CNRS, UGA, F-38000 Grenoble, France}
\author{Gabriel Seyfarth}
\affiliation{LNCMI-EMFL, CNRS, UGA, F-38000 Grenoble, France}
\author{Jacques Flouquet}
\affiliation{University Grenoble Alpes, INAC-PHELIQS, F-38000 Grenoble, France}
\affiliation{CEA, INAC-PHELIQS, F-38000 Grenoble, France}
\author{Georg Knebel}
\email{georg.knebel@cea.fr}
\affiliation{University Grenoble Alpes, INAC-PHELIQS, F-38000 Grenoble, France}
\affiliation{CEA, INAC-PHELIQS, F-38000 Grenoble, France}

\date{\today}

\begin{abstract}

We present high field magnetoresistance, Hall effect and thermopower measurements in the Ising-type ferromagnetic superconductor UCoGe.  Magnetic field is applied along the easy magnetization $c$ axis of the orthorhombic crystal. In the different experimental probes we observed five successive anomalies at $H \approx 4$, 9, 12, 16, and 21~T. Magnetic quantum oscillations were detected both in resistivity and thermoelectric power. At most of the anomalies, significant changes of the oscillation frequencies and the effective masses have been observed indicating successive Fermi surface instabilities induced by the strong magnetic polarization under magnetic field.

\end{abstract}


\maketitle

Lifshitz transitions (LTs) are continuous quantum phase transitions at zero temperature where the topology of the Fermi surface (FS) changes due to the variation of the Fermi energy and the band structure of a metal \cite{Lifshitz1960, Blanter1994}. They have been already studied in the sixties and can be induced by chemical doping, pressure, or strong magnetic field ($H$). 
However, only recently LTs have been proposed as the driving force to modify the ground-state  properties in strongly correlated electron systems.  \added{The interplay of a LT with magnetic quantum phase transitions in heavy fermion systems has been treated in various theoretical models (see e.g.~Refs.~\onlinecite{Yamaji2006, Schlottmann2011, Bercx2012, Hoshino2013, Benlagra2013, Kubo2015}). The influence of LTs on the appearance of  superconductivity is discussed in  cuprates \cite{Norman2010, LeBoeuf2011}, iron pnictides \cite{Liu2010, Wang2015} sulfur hydride \cite{Jarlborg2016} and also for the reentrance of superconductivity in URhGe \cite{Yelland2011}.  Finally, LTs play an important role in  topological insulators \cite{Liu2016}  or in the vortex state of $^3$He \cite{Silaev2015}.}


Usually the electronic band structure is a rather robust property of the metallic state, especially when applying magnetic fields. Only when the magnetic ground state is modified, changes of the FS may be detected. In a normal metal the Zeeman splitting induced by accessible magnetic fields is weak with respect to the Fermi energy which is usually of the order of a few eV. 
\deleted{Here we will focus on heavy-fermion compounds.} Importantly, in heavy fermion compounds \deleted{these systems} the Fermi energy scale is significantly reduced due to the hybridization of the conduction and the localized $f$ electrons. \added{Thus,} the Zeeman splitting of the flat bands crossing the Fermi level can be so strong that one of the spin-split FS sheets is continuously suppressed and undergoes a LT. In heavy-fermion systems a LT is often associated to  a change in the intersite and/or local magnetic fluctuations, see e.g.~in CeRu$_2$Si$_2$ \cite{Daou2006}, CeIn$_3$ \cite{Harrison2007, Sebastian2009}, YbRh$_2$Si$_2$ \cite{Rourke2008, Pfau2013, Pourret2013}. 

In this Letter we report on FS properties of UCoGe under magnetic field,  which orders ferromagnetically at $T_{C}=2.7~$K. Remarkably, homogeneous coexistence of ferromagnetism and heavy-fermion superconductivity is observed below $T_{sc}=0.6~$K \cite{Huy2007}. UCoGe crystallizes in an orthorhombic structure (space group $Pnma$). Besides the exceptional superconducting properties \cite{Aoki2011a} some normal state features of UCoGe are unique. 
The spontaneous magnetization in the ferromagnetic (FM) state is very small, $M_0\approx0.05\,\mu_B/{\rm U}$ with Ising moments along the $c$ axis \cite{Huy2008} and under magnetic field the  magnetization is strongly anisotropic with $M_c> M_b>M_a$. For $H \parallel c$, $M_c$ increases non-linearly with field and shows a broad kink at $H \approx 23$~T, but  even at $H\sim 50$~T, with $M_c \approx 0.65\,\mu_B/{\rm U}$ at $T = 1.5$~K, it is far from  saturation \cite{Knafo2012}.
Another striking point is the detection of  well separated anomalies in the magnetoresistance for  $H\parallel c$  \cite{Aoki2011, Steven2011, Bay2014} far above the collapse of the FM fluctuations ($H>1~$T) \cite{Hattori2012} while $M (H)$ rules out thermodynamic phase transitions under magnetic field at least down to 1.5~K \cite{Knafo2012}. 

In order to study the field dependence of the FS properties \added{in a highly polarizable heavy fermion system with small FS pockets} \deleted{and to clarify the nature of the successive anomalies,} 
we performed systematic resistivity ($\rho$), Hall effect ($\rho_{xy}$) and thermopower ($S$) experiments in UCoGe. \deleted{have been realized.} Two different samples (labelled S1 and S2) with residual resistivity ratios [$RRR = \rho (300\,{\rm K})/\rho (1\,{\rm K})$] of 105 and 36 have been prepared for experiments with electrical or heat current along the $b$ and $a$ axis respectively. Details of the experiments are given in the Supplemental Material \cite{Suppl}.

\begin{figure}[t]
\begin{center}
\includegraphics[width=0.9\linewidth]{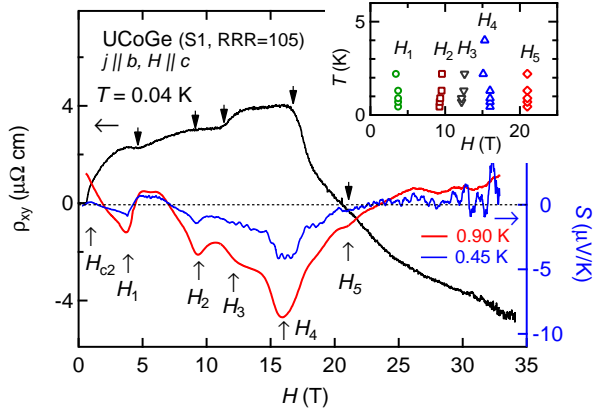}
\caption{(Color online) Hall effect $\rho_{xy}$ at 40~mK (left scale) and thermopower $S$ at 900~mK and 450~mK (right scale) of UCoGe as a function of magnetic field. A series of transitions can be observed as a function of field. The inset shows the temperature dependence of the anomalies in the thermopower.} 
\label{HallS}
\end{center}
\end{figure}

Figure \ref{HallS} shows the field dependence of $\rho_{xy}$ at 40~mK and thermopower at 900~mK and 450~mK along the $c$ axis for S1. \added{At least five} successive anomalies can be observed above the superconducting critical field $H_{c2} \approx 0.6$~T in both probes. 
\added{$S(H)$ exhibits \added{successive} marked minima at $H_1 \approx 3.65$~T, $H_2 \approx 9.2$~T, and $H_4 \approx 16$~T. A shoulder like anomaly appears at $H_3\approx 12$~T and  a small kink at $H_5\approx 21$~T. At 450~mK, in addition, large quantum oscillations occur in the thermopower (see below). At all these characteristic fields $\rho_{xy}(H)$ shows rather sharp anomalies with step-like increases or kinks. At $H_4=16~$T the most pronounced anomaly is observed and $\rho_{xy}(H)$ decreases abruptly, whereas $S(H)$ has a  marked minimum and increases for higher fields.}
\deleted{At the first anomaly $H_1\approx 3.7~$T and at the second $H_2=11~$T $\rho_{xy}(H)$ increases and shows small plateaus while $S(H)$ exhibits marked minima at $H_1 \approx 3.65$~T and $H_2 \approx 9.5$~T. The anomaly at $H_2$ is rather broad and in the following we will locate $H_2$ at the minimum of $S(H)$, while the kink in $\rho_{xy}(H)$ coincides with the maximum of $S(H)$ at 11~T. At $H_3=16~$T $\rho_{xy}(H)$ decreases abruptly whereas $S(H)$ has a marked minimum and increases for higher fields. A small kink appears at $H_4=21~$T in $S(H)$ but no clear  anomaly in $\rho_{xy}(H)$.} 
\added{In the whole field range $\rho_{xy}$ and $S$ have opposite sign,  which changes around 22$~$T suggesting a change of the dominant carrier type \cite{charge_carriers}.}  The temperature dependence of the anomalies observed in $S(H)$ is shown in the inset of Fig.~\ref{HallS}. The transitions get less pronounced with increasing temperature and disappear above $T \approx 3$~K while their field position does not change.
The clear signatures of these transitions in transport properties $\rho_{xy}(H)$ and $S(H)$ and the absence of any marked phase transition in thermodynamic properties \cite{Knafo2012, Wu2016} suggest that they are related to topological FS changes. 

\begin{figure}[t]
\begin{center}
\includegraphics[width=0.9\linewidth]{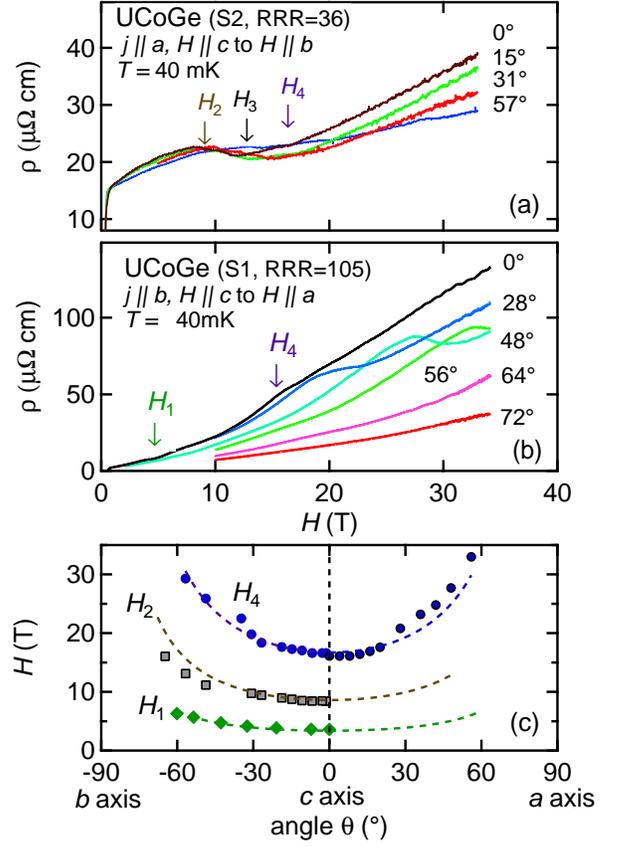}
\caption{(Color online) Transverse magnetoresistance at $T=40$~mK in UCoGe with (a) current along $a$ axis on sample S2 \deleted{(RRR=36)} for different angles in the $bc$ plane and (b) current along $b$ on S1 \deleted{(RRR=105)} in the $ac$ plane. The arrows indicate the position of the anomalies for $0^\circ$ along the $c$ axis. (c) Angular dependence of the anomalies at $H_1$, $H_2$, and $H_4$. Dashed lines are fits with $H \propto 1/\cos \theta$. } 
\label{rho}
\end{center}
\end{figure}

Figure \ref{rho} shows the transverse magnetoresistance $\rho(H)$ of UCoGe up to 34$~$T  (a) in the $bc$ plane with current along the $a$ axis (S2) and (b) in the $ac$ plane with current along $b$ (S1). The $\rho(H)$ shows in both configurations several anomalies and at high field quantum oscillations can be resolved. For $j \parallel a$ [see Fig.~\ref{rho}(a)] the resistivity shows a broad maximum around $H_2\approx9~$T \added{and a minimum at $H_3\approx 12~$T}. A tiny kink can also be observed at $H_4\approx16~$T. 
The magnetoresistance of S1 with current direction $j\parallel b$ is represented in Fig.~\ref{rho}(b) and  $\rho (H)$ increases by more than one order of magnitude between 0 and 34$~$T. Here, $\rho (H)$ is dominated by the orbital effect in the high quality sample \added{S1}. Clear anomalies at $H_1$ and $H_4$ were detected, while no clear indication of $H_2$ and $H_3$ is seen. Previously $\rho(H)$ has been reported in Ref.~\onlinecite{Bay2014} on a sample with $RRR = 30$ and current along the $b$ axis. The reported field dependence along the $c$ axis is very different from that found in the very high quality sample S1 while it is similar to that found on S2 with the current along the $a$ axis and similar $RRR$  suggesting that $\rho (H)$ is strongly sample dependent. 

In order to investigate the anisotropy of the detected anomalies we turned the samples in the $bc$ and in the $ac$ plane, while keeping the transverse configuration in both cases.  The rotation of S2 in the $bc$ plane shows a shift of the anomalies $H_2$ and $H_4$ to higher fields which can be followed up to a field angle of $\theta \approx 60^\circ$. In the $ac$ plane $\rho(H)$ is strongly reduced when the field is rotated from the easy $c$ axis to the hard $a$ axis and $H_3$ increases with angle from the $c$ axis. While the anomaly at $H_4$ smears out by rotating the field from the $c$ axis toward the $b$ axis, it gets more pronounced by rotating field towards the $a$ axis and at 48$^\circ$ a broad maximum in $\rho(H)$ appears at $H_4$. Figure \ref{rho}(c) shows the angular dependence of the anomalies in $bc$ and $ac$ planes.  The angular dependence of $H_1$ in the $bc$ plane was determined by thermopower. 
The anomalies follow quite well $1/\cos \theta$ dependence for both rotation axes and thus depend mainly on the $c$ axis component of the magnetic field. For $H_2$ good agreement with previous reports is observed \cite{Steven2011, Bay2014, Aoki2014a}. For both samples, Shubnikov-de Haas (SdH) oscillations could be observed in the magnetoresistance.

\begin{figure}[t]
\begin{center}
\includegraphics[width=0.9\linewidth]{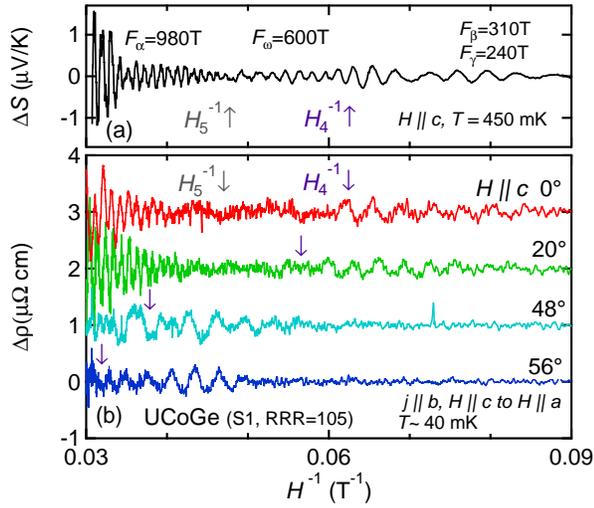}
\caption{(Color online) Quantum oscillations in UCoGe extracted from (a) thermopower and (b) resistivity  as a function of inverse magnetic field. The arrows show the positions of $H_4$ and $H_5$ anomalies detected in transport measurements. The lower panel (b) shows also quantum oscillations in the $ac$ plane measured by resistivity.} 
\label{osclncmi}
\end{center}
\end{figure}

Figure \ref{osclncmi} shows the oscillatory part after subtraction of a non-oscillatory background (see Supplemental Material) of (a) the thermopower  and (b) of the magnetoresistance of S1 for different angles in the $ac$ plane. For $H < 16$~T slow oscillations were observed
with two very close frequencies at 240 T and 310 T. These
low frequencies vanish at $H_4$ and faster oscillations with a frequency of $F_\omega = 600$~T appear  above $H_4=16$~T but disappear again at $H_5= 21$~T in the thermopower. No SdH oscillations were observed between $H_4$ and $H_5$. Above $H_5$ a higher frequency  $F_\alpha$=970$~$T called $\alpha$ branch occurs in both probes and it corresponds to that previously reported in Ref.~\onlinecite{Aoki2014a}. 

The frequency of the quantum oscillations and the corresponding effective masses in the different field intervals are reported in Table 1 of the Supplemental Material. Figure~\ref{osclncmi} (b) shows SdH oscillations at different angles in the $ac$ plane. While $H_3$ increases to higher field when approaching the $a$ axis [see  Fig.~\ref{rho}(c)], the oscillations at  $F_\gamma$ and $F_\beta$ are suppressed at $H_4$ at each angle. 
At 56$^\circ$ a continuous increase of $F_\gamma$ with field can be observed, when field gets close to the anomaly $H_4 (56^\circ)=33~$T. This suggests  that the FS pocket of the $\gamma$ branch shrinks continuously, when the field gets close to the FS reconstruction field $H_4 (56^\circ)=33~$T. Such a continuous change of a quantum oscillation frequency could not be observed clearly  for the other field directions. 
\begin{figure}[t]
\begin{center}
\includegraphics[width=0.9\linewidth]{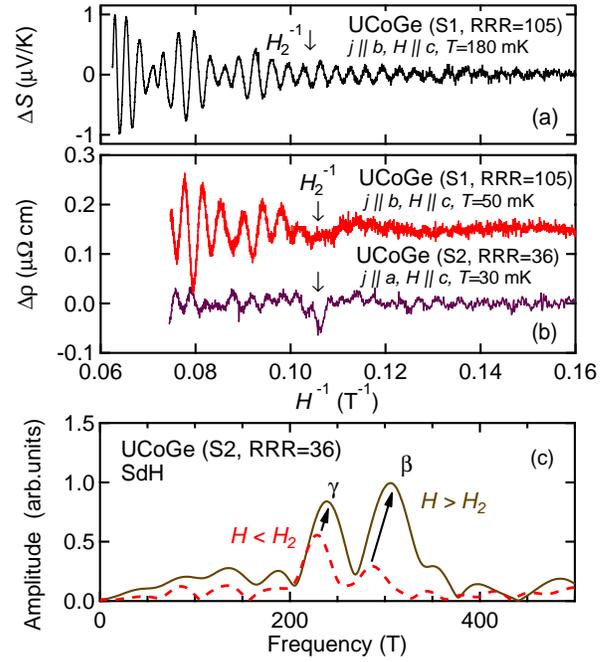}
\caption{(Color online) Quantum oscillations below 16T as a function of inverse magnetic field for (a) thermopower and (b) resistivity  measured more precisely in a superconducting magnet. (c) FFT spectrum of quantum oscillations in the resistivity of sample S2 for field along $c$ axis below and above $H_2\approx 9$~T. }  
\label{osckel}
\end{center}
\end{figure}

\begin{figure}[!t]
\begin{center}
\includegraphics[width=0.8\linewidth]{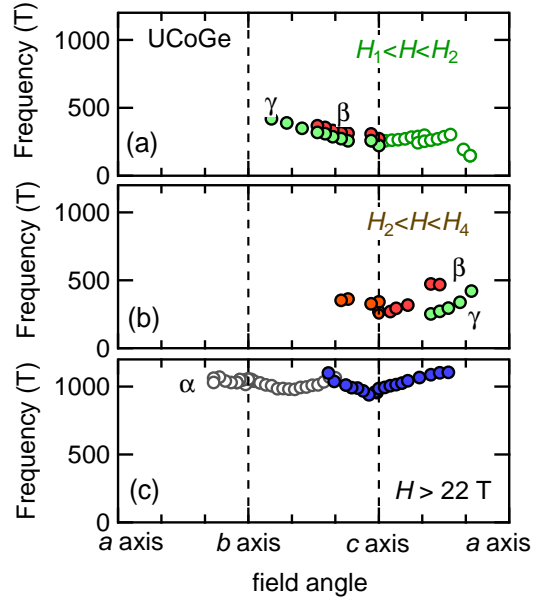}
\caption{(Color online) Angular dependence of quantum oscillation frequencies in UCoGe for the different field intervals delimited by the anomalies observed in transport measurements.  Open circles in (a) and (c) have been taken from Ref.~\onlinecite{Aoki2014a}.}
\label{angdep}
\end{center}
\end{figure}

Quantum oscillations below $H_4$=16$~$T are represented in Fig.~\ref{osckel}. Above $H_2$, a modulation of the amplitude of the oscillations in thermopower can be observed due to beating of two close quantum oscillation frequencies $F_\beta$ and $F_\gamma$. While S1 shows large oscillations above $H_2$, the SdH oscillations below 10$~$T are more visible on S2. The fast Fourier transformation (FFT) spectra of the oscillations for S2 are represented in Fig.~\ref{osckel}(c), both for field below and above $H_2$. Two frequencies can be observed below $H_2$ at 230$~$T and 280$~$T. For $H>H_2$ these two frequencies are shifted to 240$~$T and 310$~$T.  A previous dHvA study suggested a splitting of one frequency from below to above $H_2$ \cite{Aoki2014}. On the contrary, our measurements show  that both quantum oscillation frequencies survive below $H_2$ within the resolution of the FFT. Thus  a small  abrupt change in the size of the FS is directly observed by quantum oscillations at the anomaly $H_2=9~$T.

The angular dependence of the oscillation frequencies for the different field intervals are represented in Fig.~\ref{angdep}. Data in the vicinity of the $b$ axis are taken from Ref.~\onlinecite{Aoki2014} and connect perfectly to those presented here. At low field $H < H_2$, two small FS pockets elongated along the $c$ axis (ellipsoidal or cylindrical) exist. These pockets change in size at $H_2$, but disappear abruptly above $H_4$. The angular dependence of the frequency at $F_\omega = 600$~T  has not been measured. The pocket $\alpha$ with the heavy effective mass ranging from 17 to 23~$m_0$   seems to be nearly spherical with a frequency around $F_\alpha \approx 1000~$T [Fig.~\ref{angdep}(c)] \added{ and is experimentally observed above 22~T, independent of the field angle.} \deleted{ From the angular dependence of the anomalies in the transport [see Fig.~\ref{rho}(c)] we can conclude that this frequency also exists below $H_4$. }

The main observation is that \added{most} \deleted{all} anomalies observed in the field dependence of the transport properties (see Figs.~\ref{HallS} and \ref{rho}) coincide with abrupt changes in the quantum oscillation frequencies and effective masses (see Table 1 in Supplemental Material). They are related to modifications of the FS topology with the most drastic change occurring at $H_4$ where the Hall effect collapses and $S(H)$ has a pronounced minimum. The FS can be easily modified by applying a magnetic field and the small FS pockets disappear through a LT. We can estimate the characteristic energy  of each detected pocket with $\epsilon_i=\hbar^2k_{F,i}^2/2m^\star_i \approx \hbar eF_i/m_i^\star c$ and we find $\epsilon_\gamma \approx 2.5$~meV, $\epsilon_\omega \approx 5$~meV and $\epsilon_\alpha \approx 6.6$~meV. These energies can be compared to the  Zeeman energy scale of a free electron divided by field, $\epsilon_Z/\mu_0H = g\mu_B \approx 0.12$~meV/T for $g = 2$. As UCoGe is a weak ferromagnet this effect will even be  strengthened by the internal field. 
Hence an  important polarization of the bands can be achieved by \deleted{experimentally} easily accessible magnetic fields and thus a series of magnetic field-induced LT appears.  

The magnetization up to 50$~$T \cite{Knafo2012} has a strongly non-linear field dependence suggesting that the electronic magnetic response must vary strongly with the magnetic field while the FM fluctuations are already fully suppressed for $H>1$~T along the $c$ axis \cite{Aoki2011a, Hattori2012}.  
Thus the electronic instabilities seem to occur in the paramagnetic regime without any additional phase transitions and far above the field where the FM intersite magnetic correlations collapse. 
The key phenomenon is that FS changes are induced by crossing some critical values of magnetic polarization.  \deleted{Sometimes}\added{In some systems} such FS changes are accompanied by a metamagnetic-like transition depending on the nature of the electronic instability.
Very recent magnetization measurements \cite{Nakamura} point to a tiny metamagnetic-like transition at $H_2$ but do not detect any anomaly at $H_1$ and the high field measurements \cite{Knafo2012} exclude it for $H_4$ and $H_5$. The case of UCoGe can be compared to the series of \deleted{LTs} \added{FS reconstructions} observed inside the hidden order phase of URu$_2$Si$_2$.  In  this compound no detectable effects on the bulk magnetization have been observed \cite{Scheerer2012}, but the LTs are related to the polarization of the small FS pockets \cite{Altarawneh2011, Malone2012, Pourret2013}. In CeRu$_2$Si$_2$ the LT is linked to the pseudo-metamagnetic transition where one spin-split FS vanishes continuously at the transition \cite{Daou2006, Boukahil2014}, while in YbRh$_2$Si$_2$ the LT \cite{Rourke2008, Pourret2013a} goes along with a suppression of the local Kondo effect as has been demonstrated by renormalized band structure calculations under magnetic field \cite{Pfau2013}. Recently, a LT occurring at 28~T  has been reported in paramagnetic CeIrIn$_5$ \cite{Aoki2016}.

Different LDA band structure calculations have been performed on UCoGe \cite{Samsel-Czekala2010, Fujimori2015}  showing strong differences in the FS topology. In the paramagnetic state three bands are contributing to FS sheets with rather small volume, characteristic for a low carrier or semimetallic system. Two  cigar-like \cite{Samsel-Czekala2010} or pillar-like \cite{Fujimori2015} \added{electron} FSs centered around the $S$ point have been predicted which may correspond to the small FSs observed below $H_2$. 
In Ref.~\onlinecite{Samsel-Czekala2010} the FM state with a magnetic polarization along the $c$ axis has also been calculated. The FSs (with a moment of $-0.47\mu_B/{\rm U}$ much larger than in experiment) differ significantly from the paramagnetic ones and do not at all agree with our experiment. In ARPES experiments at zero field details of the FS could not be resolved up to now \cite{Fujimori2015} and it will be of great interest to see the differences of the FS above and below $T_C$.  

In conclusion, we give clear evidences by quantum oscillation experiments for FS instabilities under magnetic field in UCoGe for field along the easy magnetization axis. 
\added{The occurrence of several LTs under field in the polarized phase of UCoGe shows that FS properties of heavy fermion systems  can be easily tuned by magnetic field. The LTs are decoupled from intersite correlations and seem to be driven only by changes in the local fluctuations  induced by reaching a critical magnetic polarization.
It unveils a strong interplay between magnetic polarization and FS topology, which is directly linked with the dual localized and itinerant nature of the 5$f$ electrons. A key challenge in theory is now to take into account the feedback between polarization and FS to model the influence of the magnetic field on the electronic structure. }


\begin{acknowledgments}
We thank J.-P. Brison, B. Wu, V.~P.~Mineev, A.~A.~Varlamov and G.~Zwicknagl for fruitful discussions.
This work has been supported by the ERC grant "NewHeavyFermion", and  KAKENHI (25247055,15H05884,15H05882,15K21732, 16H04006), the EuromagNET II (EU contract no. 228043), LNCMI-CNRS is member of the European Magnetic Field Laboratory (EMFL).
\end{acknowledgments}
%
%

%

\section{Supplemental Material}

\subsection{Single crystal growth}
High quality single crystals of UCoGe (orthorhombic TiNiSi structure) were grown in a tetra-arc furnace by the Chzochralski method and further annealed. Two single crystals have been cut by spark erosion. The orientation of the single crystals have been checked by Laue diffraction. Sample S1 has a  residual resistivity ratio [$RRR = \rho (300\,{\rm K})/\rho (1\,{\rm K})$] of 105 and  sample S2 of 36. 

\subsection{Experimental details}
Resistivity measurements were performed under high magnetic field in a 34$~$T resistive magnet at the high magnetic field facilility LNCMI Grenoble. The samples were mounted on a one axis mechanical rotator and cooled in a top loading dilution refrigerator down to 40~mK.  The resistivity was measured with the usual four probe ac current method. The electrical current was below 200$~\mu$A. More accurate measurements were performed in CEA in a 13$~$T superconducting magnet. Sensitivity was enhanced compared to the high field set up by the use of a low temperature transformer thermalized at the 1~K stage of the dilution refrigerator. These measurements were performed down to $30~$mK and electrical current was reduced down to 50$~\mu$A. 

Thermoelectric power measurements under high magnetic field were performed in LNCMI Grenoble, with a $^3$He cryostat down to 400$~$mK and with a resistive magnet up to 34$~$T. Lower temperature measurements were done in a superconducting magnet up to 16$~$T and with a home-made dilution fridge down to 0.1$~$K. To determine the angular dependence of $H_1$ in the $bc$-plane an Attocube piezorotator has  been used to rotate the sample in the magnetic field.

The thermopower has been measured with an "one heater two thermometers" set up. The RuO$_2$ thermometers had been calibrated against a Ge thermometer which is installed in the field-compensated zone of the superconducting magnet. Both thermometers and the heater are thermally decoupled from the heat bath of the sample holder by highly resistive manganine wires with specific resistance of 200~$\Omega/$m. To measure quantum oscillations the field has been swept continuously upwards. A constant power has been used to heat the sample. The thermoelectric voltage had been taken at the beginning and the end of the field sweeps to estimate the background signal. 

\subsection{Quantum oscillations}

To extract the oscillating part of the magnetoresistance (see Fig.~2 of the article), a Loess algorithm has been used to average the oscillations and the smoothed curve has been subtracted from the data. As in the thermoelectric signal the different anomalies are very marked and the signal is very non-monotonous, we subtracted the data at $T=1.2$~K where no oscillations but still the anomalies are visible.  Then, we determined the envelope of the remaining signal and subtracted the average of the upper and lower envelope. This was cross-checked by subtracting a polynomial background on a reduced window. The quantum oscillation frequencies have been determined by fast Fourier transformation (FFT) in the respective field windows between the different anomalies. The change of the FFT frequencies has been also confirmed by performing a FFT on sliding windows in $1/H$.

Table \ref{table} gives the quantum oscillation frequencies and the effective masses for field along the $c$ axis in the respective field intervals between the different anomalies.  While the temperature dependence of the amplitude of the SdH oscillations follow the Lifshitz Kosevich formula, the amplitude of the thermoelectric power oscillations is given by the derivative of the Lifshitz Kosevich formula \cite{Pantsulaya1989, PalacioMorales2016}. A very good agreement was observed between quantum oscillation  frequencies observed on the different samples with resistivity and with thermopower on S1 and a relatively good agreement for the effective masses.

\begin{table*}[t]
\begin{ruledtabular}
	
	\caption{\label{table}Quantum oscillations frequencies and effective masses in UCoGe from resistivity and Seebeck effect measurements for field along the $c$ axis. Different field intervals are considered, they are delimited by the anomalies observed in transport measurements.}
	
		\begin{tabular}{cccccccc}
		\hline
	&	&  \multicolumn{2}{c} {SdH(sample1)} & \multicolumn{2}{c} {SdH(sample2)} & \multicolumn{2}{c} {Seebeck(sample1)} \\
	$H$ range	&	orbit  & $F$(T) & $m^*(m_0)$ & $F$(T) & $m^*(m_0)$ & $F$(T) & $m^*(m_0)$ \\
		\hline
	$4{\rm T}<H<9{\rm T}$	&$\gamma$ &  &  & 230 & 7 &  &\\
	&	$\beta$ &270 & & 280 & & 285 &	\\
	$9{\rm T}<H<16{\rm T}$&	$\gamma$ & 240 & 11 & 240 & 8 & 240 & 12\\
	&	$\beta$ & 310 & 11 & 310 & 11 & 310 &	13\\	
	$16{\rm T}<H<21{\rm T}$&	$\omega$ &   & & & &600 & 14 \\	
	$23{\rm T}<H$&	$\alpha$ & 970 & 17 & 955 & &980 & 14 \\
		\end{tabular}
	
\end{ruledtabular}
\end{table*}

\end{document}